\documentclass[preprint,5p,times,twocolumn]{elsarticle}

\usepackage{amsmath,graphicx,verbatim,epsfig}
\usepackage[utf8x]{inputenc}
\usepackage{slashed}
\usepackage{amssymb}
\usepackage{bm}
\usepackage[colorlinks=true,linkcolor=blue,citecolor=blue]{hyperref}

\newcommand{\Ds}{\displaystyle}

\newcommand{\nn}{\nonumber}

\renewcommand{\(}{\left(}
\renewcommand{\)}{\right)}
\renewcommand{\[}{\left[}
\renewcommand{\]}{\right]}

\newcommand{\Tr}{\text{Tr}}

\newcommand{\ot}{\leftarrow}

\renewcommand{\vec}[1]{\bm{#1}}

\begin{document}

% To change the format, if desired...
%\baselineskip 3.0ex
%\vspace*{18pt}

\title{Twist-2 matching of transverse momentum dependent distributions}

%%%%%%%%%%%%%%%%%
%Headings for revtex4 package
%%%%%%%%%%%%%%%%%

\author[Madrid]{Daniel Guti\'errez-Reyes}
\ead{dangut01@ucm.es}

\author[Madrid]{Ignazio Scimemi}
\ead{ignazios@fis.ucm.es}

\author[Regensburg]{Alexey A. Vladimirov}
\ead{aleksey.vladimirov@gmail.com}

\address[Madrid]{Departamento de F\'isica Te\'orica II,
Universidad Complutense de Madrid (UCM),\\
28040 Madrid, Spain}

\address[Regensburg]{Institut f\"ur Theoretische Physik, Universit\"at Regensburg,\\
D-93040 Regensburg, Germany}

%\date{\today, \currenttime}
%%%%%%%%%%%%%%%%%%%%%%%%%%%%%
%%%%%%%%%%%%%%%%%%%%%%%%%%%%%

% For revtex4:
\begin{abstract}

We systematically study the large-$q_T$ (or small-$b$) matching of transverse momentum dependent (TMD) distributions to the twist-2 integrated parton distributions. Performing operator product expansion for a generic TMD operator at the next-to-leading order (NLO) we found the complete set of TMD distributions that match twist-2. These are unpolarized, helicity, transversity, pretzelosity and linearly polarized gluon distributions. The NLO matching coefficients for these distributions are presented. The pretzelosity matching coefficient is zero at the presented order, however, it is evident that it is non-zero in the following orders. This result offers a natural explanation of the small value of pretzelosity found in phenomenological fits. We also demonstrate that the cancellation of rapidity divergences by the leading order soft factor imposes the necessary requirement on the Lorentz structure of TMD operators, which is supported only by the TMD distributions of leading dynamical twist. Additionally, this requirement puts restrictions on the $\gamma_5$-definition in the dimensional regularization.
% For revtex4:
\end{abstract}

% Add this command for revtex4, not for jheppub.
\maketitle

%Add these lines for jheppub
%\begin{document}
%\maketitle

%%%%%%%%%%%%%%%%%%%%%%%%%%%%%%%%%
%%%%%%%%%%%%%%%%%%%%%%%%%%%%%%%%%
%%%%%%%%%%%%%%%%%%%%%%%%%%%%%%%%%
\section{Introduction}
%%%%%%%%%%%%%%%%%%%%%%%%%%%%%%%%%
The transverse momentum dependent (TMD) factorization theorems for semi-inclusive deep inelastic scattering (SIDIS) and Drell-Yan type processes formulated in  \cite{GarciaEchevarria:2011rb,Collins:2011zzd,Chiu:2012ir,Echevarria:2014rua} allow a  consistent treatment of rapidity divergences in the definition of spin (in)dependent TMD distributions. They also provide a self-contained definition of TMD operators which can be considered individually by standard methods of quantum field theory without referring to a scattering process. In particular, the large-$q_T$ (or small-$b$) matching of TMD distributions on the corresponding integrated functions can be evaluated. Such consideration is practically very important because the resulting matching coefficients serve as an initial input to many models and phenomenological ansatzes for TMD distributions. The unpolarized TMD distribution is the most studied case and it has been treated using different regularization schemes at the next-to-leading order (NLO) \cite{GarciaEchevarria:2011rb,Becher:2010tm,Collins:2011zzd,Echevarria:2014rua,Aybat:2011zv,Cherednikov:2008ua,Cherednikov:2007tw} and the next-to-next-to-leading order (NNLO)~\cite{Echevarria:2015usa,Echevarria:2016scs,Gehrmann:2014yya,Gehrmann:2012ze}. For polarized distributions such a program has been performed only for helicity, transversity and linearly polarized distributions at NLO \cite{Bacchetta:2013pqa,Echevarria:2015uaa}. However, these works miss a systematic discussion on the relevant renormalization schemes, which are fundamental to establish their calculation and to provide a spring to higher order analysis. By this article we open a series of articles devoted to the study of the small-$b$ matching of polarized TMD distributions. The primary goal of this letter is to provide a dedicated and consistent study of the leading twist (twist-2) matching of the TMD operators.
 
The quark and gluon components of the generic TMD operators are
\begin{align}\label{def:TMD_OP_Q}
\Phi_{ij}(x,\vec b)&=\int \frac{d\lambda}{2\pi}e^{-ixp^+\lambda}\bar q_i\(\lambda n+\vec b\)\mathcal{W}(\lambda,\vec b)q_j\(0\),
\\ \label{def:TMD_OP_G}
\Phi_{\mu\nu}(x,\vec b)&=\frac{1}{xp^+}\int \frac{d\lambda}{2\pi}e^{-ixp^+\lambda}
F_{+\mu}\(\lambda n+\vec b\)\mathcal{W}(\lambda,\vec b)F_{+\nu}\(0\),
\end{align}
where $n$ is the lightlike vector and we use the standard notation for the lightcone components of vector $v^\mu=n^\mu v^-+\bar n^\mu v^++g_T^{\mu\nu}v_\nu$ (with $n^2=\bar n^2=0$, $n\cdot\bar n=1$, and $g_T^{\mu\nu}=g_{\mu\nu}-n^\mu \bar n^\nu-\bar n^\mu n^\nu$). The operator $\mathcal{W}$ is
\begin{eqnarray}
\mathcal{W}(\lambda,\vec b)=\tilde W_{n}^T\(\lambda n+\vec b\)\sum_X |X\rangle\langle X|\tilde W_{n}^{T\dagger}\(0\),
\end{eqnarray}
with Wilson lines $W$ taken in the appropriate representation of gauge group . The staple contour of the gauge link results in the rapidity divergences, the unique feature of TMD operators. The rapidity divergences are removed by the proper rapidity renormalization factor $R$, which is built from the TMD soft factor,
\begin{align} \label{eq:softf}
S(\vec b) &=
\frac{{\Tr}_\text{color}}{N_c}
\langle 0|\[S_n^{T\dagger} \tilde S_{\bar n}^T \](\vec b)
\[\tilde S^{T\dagger}_{\bar n} S_n^T\](0)|0\rangle,
\end{align}
where $S_n$ and ${\tilde S}_{\bar n}$ stand for soft Wilson lines along $n$ and $\bar n$ (for the precise definition of $W^T$ and $\tilde S^T$ see e.g.\cite{Echevarria:2016scs}). The structure of factor $R$ follows from the TMD factorization theorem \cite{GarciaEchevarria:2011rb,Collins:2011zzd,Echevarria:2014rua,Echevarria:2016scs,Chiu:2012ir} and  depends on the rapidity regularization scheme. However, the expressions for rapidity-divergence-free quantities, such as evolution kernels and matching coefficients are independent on the scheme. In the following we use the $\delta$-regularization scheme formulated in \cite{Echevarria:2015byo,Echevarria:2016scs}. This scheme uses the infinitesimal parameter $\delta$ as a regulator for rapidity divergences in combination with  the usual dimensional regularization (with $d=4-2\epsilon$, $\epsilon>0$) for ultraviolet and collinear divergences. Such combination appears to be very visual and practically convenient. The central statement of the TMD factorization theorem is the complete elimination of rapidity divergences by the rapidity renormalization factor $R$. In the $\delta$-regularization scheme where $R=1/\sqrt{S(\vec b)}$, the rapidity divergences take the form of $\ln\delta$ and do not mix with divergences in $\epsilon$, which yield the exact cancellation of $\ln\delta$ at finite $\epsilon$. This non-trivial demand is necessary for a consistent higher-then-NLO evaluation and requires the matching of regularizations for different field modes (see \cite{Echevarria:2016scs}). It also results into the correspondence between TMD processes and the jet production \cite{Vladimirov:2016dll}.

The hadron matrix elements of the TMD operators with open vector and spinor indices (\ref{def:TMD_OP_Q},\ref{def:TMD_OP_G}) are to be decomposed over all possible Lorentz variants, which define TMD parton distribution functions (TMDPDFs). In the literature, this decomposition has been made in the momentum space (for spin-1/2 hadrons it can be found in \cite{Goeke:2005hb,Bacchetta:2008xw} (for quark operators) and in \cite{Mulders:2000sh} (for gluon operators)). However, it is convenient to consider TMD distributions in the impact parameter space, where it is naturally defined. The correspondence between decomposition in momentum and impact parameter spaces can be found in e.g.\cite{Boer:2011xd,Echevarria:2015uaa}. In this work we need only a part of the complete decomposition,
%\begin{widetext}
\begin{align}
\label{TMD_Q_dec}
&\Phi_{q\ot h,ij}(x,\vec b)=\langle h|\Phi_{ij}(x,\vec b)|h\rangle =\frac{1}{2}\Big(
f_1\gamma^-_{ij} +g_{1L} S_L(\gamma_5\gamma^-)_{ij}
\\\nn&+(S_T^\mu
i\gamma_5\sigma^{+\mu})_{ij}h_1+(i\gamma_5\sigma^{+\mu})_{ij}\(\frac{g_T^{\mu\nu}}{2}+\frac{ b^\mu  b^\nu}{ \vec b^2}\)\frac{S_{T}^\nu }{2}h_{1T}^{\perp}+...\Big),
\\\label{TMD_G_dec}
&\Phi_{g\ot h,\mu\nu}(x,\vec b)=\langle h|\Phi_{\mu\nu}(x,\vec b)|h\rangle
\\\nn&
= \frac{1}{2}\Big(-g_T^{\mu\nu}f_1^g -
i\epsilon_T^{\mu\nu}S_Lg_{1L}^g +2h_1^{\perp g} \(\frac{g_T^{\mu\nu}}{2}+\frac{ b^\mu  b^\nu}{\vec b^2}\)+...\Big),
\end{align}
%\end{widetext}
where  the vector $b^\mu$ is a 4-dimensional vector of the impact parameter ($b^+=b^-=0$ and $ -b^2\equiv \vec b^2>0$), and $S_{T,L}$ are components of the hadron spin vector defined in Eq.(\ref{def:spin}). On the r.h.s. of  Eqs.~(\ref{TMD_Q_dec},\ref{TMD_G_dec})  and in the rest of the letter we omit arguments of TMD distributions $(x,\vec b)$, unless they are necessary. Note that in Eq.~(\ref{TMD_Q_dec}) we use the normalization for the distribution $h_{1T}^\perp$ different from the traditional one \cite{Goeke:2005hb}. The traditional definition can be recovered substituting $h_{1T}^{\perp}\rightarrow h_{1T}^{\perp} \vec b^2 M^2$, with $M$ being the mass of hadron. In the section \ref{sec:trans_and_pretz} we argue that such normalization is natural.

In Eqs.(\ref{TMD_Q_dec},\ref{TMD_G_dec}) we write only the TMD distributions that match the twist-2 integrated distributions. The dots include the TMD distributions that match the twist-3 and higher parton distribution functions (PDF). The reported distributions are usually addressed as  helicity ($g_{1L}$ and $g_{1L}^g$), transversity ($h_1$), pretzelosity ($h_{1T}^\perp$) and linearly polarized gluon ($h_1^{\perp g}$) distributions. The small-$b$ matching of these distributions has been performed separately for quarks \cite{Bacchetta:2013pqa} and gluons \cite{Echevarria:2015uaa} in different renormalization schemes. Furthermore, the pretzelosity distribution has been overlooked by these groups. In this letter, we present a uniform and consistent  NLO matching of these TMD distributions.

%%%%%%%%%%%%%%%%%%%%%%%%%%%%%
\section{Small-$b$ operator product expansion}
\label{sec:OPE}
%%%%%%%%%%%%%%%%%%%%%%%%%%%%5

The small-$b$ operator product expansion (OPE) is the relation between TMD operators and lightcone operators. Its leading order can be written as
\begin{align}\label{OPE_Q}
&\Phi_{ij}(x,\vec b)=
\\\nn&\quad
\Big[\(C_{q\ot q}(\vec b)\)_{ij}^{ab}\otimes \phi_{ab}\Big](x)+\Big[\(C_{q\ot g}(\vec b)\)_{ij}^{\alpha\beta}\otimes
\phi_{\alpha\beta}\Big](x)+...,
\\\label{OPE_G}
&\Phi_{\mu\nu}(x,\vec b)=
\\\nn&\quad\Big[\(C_{g\ot q}(\vec b)\)_{\mu\nu}^{ab}\otimes \phi_{ab}\Big](x)+\Big[\(C_{g\ot g}(\vec b)\)_{\mu\nu}^{\alpha\beta}\otimes
\phi_{\alpha\beta}\Big](x)+...,
\end{align}
where symbol $\otimes$ denotes the Mellin convolution in the variable $x$. The functions $C(\vec b)$ are dimensionless, i.e. they depend on $\vec b$ only logarithmically. The dots represent the power suppressed contributions, which presently  have been studied only for the unpolarized case (see discussion in \cite{Scimemi:2016ffw}). At this order of OPE, the functions $\phi(x)$ are the formal limit of the TMD operators $\Phi(x,\vec 0)$. The hadronic matrix elements of $\phi$ are the PDFs
\begin{eqnarray}\label{PDF_Q_dec}
\phi_{q\ot h,ij}(x)&=&\langle h|\phi_{ij}(x)|h\rangle \\\nn &=&\frac{1}{2}\Big(
f_q(x)\gamma^-_{ij} +\Delta f_q(x) S_L(\gamma_5\gamma^-)_{ij}
\\\nn&&+(S_T^\mu
i\gamma_5\sigma^{+\mu})_{ij}\delta f_q(x)\Big)+\mathcal{O}\(\frac{M}{p^+}\),
\\\label{PDF_G_dec}
\phi_{g\ot h,\mu\nu}(x)&=&\langle h|\phi_{\mu\nu}(x)|h\rangle 
\\\nn &=& \frac{1}{2}\(-g_T^{\mu\nu}f_g(x) -
i\epsilon_T^{\mu\nu}S_L\Delta f_g\)+\mathcal{O}\(\frac{M}{p^+}\),
\end{eqnarray}
where $M$ is the mass of hadron, $S_{L}$ and $S_T$ are the components of the hadron spin vector 
\begin{eqnarray}\label{def:spin}
S^\mu=S_L\(\frac{p^+}{M}\bar n^\mu-\frac{M}{2p^+}n^\mu\)+S_T^\mu ,
\end{eqnarray}
and $\epsilon_T^{\mu\nu}=\epsilon^{+-\mu\nu}=n_\alpha \bar n_\beta \epsilon^{\alpha\beta\mu\nu}$. For future convenience we introduce the universal notation
\begin{eqnarray}
\Phi_q^{[\Gamma]}=\frac{\Tr(\Gamma \Phi)}{2},\qquad \Phi_g^{[\Gamma]}=\Gamma^{\mu\nu}\Phi_{\mu\nu}.
\end{eqnarray}
Both sides of Eq.~(\ref{OPE_Q},\ref{OPE_G}) should be supplemented by the ultraviolet renormalization constants. Additionally, the TMD operator on the l.h.s. is to be multiplied by the rapidity renormalization factor $R$. The renormalized TMD operator has the form 
\begin{align}\label{TMD_ren}
\Phi^{\text{ren}}\!(x,\vec b;\mu,\zeta)
=Z(\mu,\zeta|\epsilon)R(\vec b,\mu,\zeta|\epsilon,\delta)\Phi(x,\vec b|\epsilon,\delta),
\end{align}
where we explicitly show the dependence on regularization parameters on the r.h.s.. The dependence on $\epsilon$ and $\delta$ cancels in the product. The renormalization factors are independent on the Lorentz structure but dependent on parton flavor. The explicit expressions for these factors up to NNLO can be found in \cite{Echevarria:2016scs,Echevarria:2015byo}.

The cancellation of rapidity  divergences for the spin-dependent distributions is a non-trivial statement.  Let us consider the small-$b$ OPE for a generic TMD quark operator. At one loop  we find
%\begin{widetext}
\begin{align}\label{gen_OPE}
&\Phi_q^{[\Gamma]}=\Gamma^{ab}\phi_{ab}+a_sC_F\pmb B^{\epsilon}\Gamma(-\epsilon)\Big[
\\\nn&
-(\gamma^+\gamma^-\Gamma+\Gamma\gamma^-\gamma^+)^{ab}
+\bar x\(\frac{g_T^{\alpha\beta}}{2}-\frac{ b^\alpha  b^\beta}{4\pmb B}\epsilon\)(\gamma^\mu \gamma_\alpha \Gamma \gamma_\beta \gamma_\mu)^{ab}
\\&\nn +\(\frac{1}{(1-x)_+}-\ln\(\frac{\delta}{p^+}\)\)\(\gamma^+\gamma^-\Gamma+\Gamma\gamma^-\gamma^+ +
\frac{i\epsilon \gamma^+\!\!\!\not  b\Gamma}{2\pmb B}+\frac{i\epsilon \Gamma\!\!\!\not  b\gamma^+}{2\pmb B}\)^{ab}
\\&\nn
-\frac{i\pi}{2}\(\gamma^+\gamma^-\Gamma-\Gamma\gamma^-\gamma^+ +
\frac{i\epsilon \gamma^+\!\!\!\not b\Gamma}{2\pmb B}-\frac{i\epsilon \Gamma\!\!\!\not  b\gamma^+}{2\pmb B}\)^{ab}
\Big]\otimes \phi_{ab}+\mathcal{O}(a_s^2),
\end{align}
%\end{widetext}
where $\pmb B= \vec b^2/4>0$, $a_s=g^2/(4\pi)^{d/2}$, and we use the standard PDF notation, $[f(x)]_+=f(x)-\delta(\bar x)\int dy f(y)$ and $\bar x=1-x$. In this expression, we omit the gluon operator contribution for simplicity. The complex term in the last line of Eq.~(\ref{gen_OPE}) is the artifact of $\delta$-regularization. The logarithm of $\delta$ represents the rapidity divergence which is to be eliminated by the factor $R$ which at this perturbative order reads
\begin{align}
R&=1+
2a_sC_F\pmb B^\epsilon\Gamma(-\epsilon)
\nn \\
&\times \Big(
\mathbf{L}_{\sqrt{\zeta}}+2\ln\(\frac{\delta}{p^+}\)-\psi(-\epsilon)-\gamma_E\Big)+\mathcal{O}(a_s^2),
\end{align}
where  $\mathbf{L}_X=\ln\(\pmb B X^2 e^{2\gamma_E}\)$. The rapidity divergence cancels in the product $R\Phi$ if and only if
\begin{eqnarray}\label{gG=0}
\gamma^+\Gamma=\Gamma \gamma^+=0\ ,
\end{eqnarray}
yielding 
\begin{eqnarray}\label{eq:OPE_qq}
&&R\Phi_q^{[\Gamma]}=\Gamma^{ab}\phi_{ab}+a_sC_F\pmb B^{\epsilon}\Gamma(-\epsilon)
\\\nn &&\times \Big[
\(-4+\frac{4}{(1-x)_+}+2\delta(\bar x)(\mathbf{L}_{\sqrt{\zeta}}-\psi(-\epsilon)-\gamma_E)\)\Gamma^{ab}
\\\nn&&
+\bar x\(\frac{g_T^{\alpha\beta}}{2}-\frac{ b^\alpha  b^\beta}{4\pmb B}\epsilon\)(\gamma^\mu \gamma_\alpha \Gamma \gamma_\beta \gamma_\mu)^{ab}
\Big]\otimes \phi_{ab}+\mathcal{O}(a_s^2)\ .
\end{eqnarray}
The cancellation of rapidity divergences is the fundamental pre-requisite to obtain the matching coefficients of the renormalized operator  $\Phi$ and $\phi$.

The conditions analogue to Eq.~(\ref{gG=0}) for the gluon operator are 
\begin{eqnarray}\label{Lambda+=0}
 \Gamma^{+\mu}=\Gamma^{-\mu}=\Gamma^{\mu+}=\Gamma^{\mu-}=0.
\end{eqnarray}
They follow from OPE for a generic gluon TMD operator $\Phi^{\mu\nu}$ similar to Eq.~(\ref{gen_OPE}), which we do not present here, since it is rather lengthy and not instructive. The conditions in Eq.~(\ref{gG=0},\ref{Lambda+=0}) are satisfied only for the following Lorentz structures
\begin{align}
\Gamma^q=\{\gamma^+,\gamma^+\gamma^5,\sigma^{+\mu}\},&&
\Gamma^g=\{g_T^{\mu\nu},\epsilon_T^{\mu\nu}, b^\mu  b^\nu/\vec b^2\},
\end{align}
which exactly correspond to the Lorentz structures for the so called   \lq\lq{}leading dynamical twist\rq\rq{} TMD distributions. In this way, the relations Eq.~(\ref{gG=0},\ref{Lambda+=0}) provide a definition of the leading dynamical twist for TMD operators that can be used with no reference to a particular cross-section.  On the other hand, our consideration shows that TMD operators of non-leading dynamical twist have rapidity singularities that are not canceled by the soft factor in Eq.~(\ref{eq:softf}).  While  we have no knowledge of a calculation of the correction to the leading order of TMD factorization, our finding demonstrates that it   has a different structure of rapidity divergences (which can spoil the factorization). The relation in Eq.~(\ref{gG=0}) will be used in the next section to fix the definition of $\gamma_5$ in the dimensional regularization.

In order to calculate the matching coefficients, we consider the quark and gluon matrix elements with the momentum of parton set to $p^\mu=p^+\bar n^\mu$. This choice of kinematic is allowed for consideration of twist-2 contribution only (which is the case of this article). Then, the calculations are greatly simplified. In particular, the perturbative corrections to the parton matrix element of $\phi$'s are zero, due to the absence of a scale in the dimensional regularization. Therefore, such matrix elements are equal to their renormalization constant, i.e. has not finite in $\epsilon$-terms. In practice, it implies that the matching coefficient is the $\epsilon$-finite part of the parton matrix element of the renormalized TMD operator (\ref{TMD_ren}). The evaluation of OPE for a general Lorentz structure (as in Eq.~(\ref{eq:OPE_qq})) is not very representative because one needs only the components associated with the TMDPDFs. Therefore, we project out the required components and present the expressions for each particular distribution.

%%%%%%%%%%%%%%%%%%
\section{Helicity distribution}
\label{sec:helicity}
%%%%%%%%%%%%%%%
In the case of helicity distributions the Lorentz structures for quark and gluon operators are
\begin{eqnarray}
\Gamma=\gamma^+\gamma^5 ,\quad \Gamma^{\mu\nu}=i\epsilon_T^{\mu\nu}.
\end{eqnarray}
The corresponding \lq\lq{}orthogonal\rq\rq{} projectors are
\begin{eqnarray}\label{hel:proj}
\overline{\Gamma}=\mathcal{N}_{\text{sch.}}\frac{\gamma^-\gamma^5}{2} ,\quad \overline{\Gamma}^{\mu\nu}=i\mathcal{N}_{\text{sch.}}\frac{\epsilon_{T}^{\mu\nu}}{2},
\end{eqnarray}
where the factor $\mathcal{N}_{\text{sch.}}$ depends on the definition of $\gamma_5$ matrix in dimensional regularization. Historically the most popular schemes (for QCD calculations) are 't Hooft-Veltman-Breitenlohner-Maison (HVBM) \cite{tHooft:1972tcz,Breitenlohner:1977hr}, and Larin scheme \cite{Larin:1993tq,Larin:1991tj}. In both schemes the combination $\gamma^+\gamma^5$ can be presented as
\begin{eqnarray}
\gamma^+\gamma^5=\frac{i}{3!}\epsilon^{+\nu\alpha\beta}\gamma_{\nu}\gamma_\alpha\gamma_\beta,
\end{eqnarray}
where $\epsilon^{\mu\nu\alpha\beta}$ is the antisymmetric Levi-Civita tensor. The difference between schemes is hidden in the definition of Levi-Civita tensor. In HVBM the $\epsilon^{\mu\nu\alpha\beta}$ is defined only for 4-dimensional set of indices. I.e. $\epsilon^{\mu\nu\alpha\beta}=1$ if $\{\mu\nu\alpha\beta\}$ is even permutation of $\{0,1,2,3\}$, $\epsilon^{\mu\nu\alpha\beta}=-1$ if the permutation is odd, and  $\epsilon^{\mu\nu\alpha\beta}=0$ for any another case. In Larin scheme the $\epsilon$-tensor is non-zero for all set of $d$-dimensional indices. The value of individual components are undefined, however, the product of two $\epsilon$-tensors is defined, $\epsilon^{\mu_1\nu_1\alpha_1\beta_1}\epsilon^{\mu_2\nu_2\alpha_2\beta_2}=-g^{\mu_1\mu_2}g^{\nu_1\nu_2}g^{\alpha_1\alpha_2}g^{\beta_1\beta_2}+g^{\mu_1\nu_2}g^{\nu_1\mu_2}g^{\alpha_1\alpha_2}g^{\beta_1\beta_2}-..$, where the dots mean all $4!$ permutations of indices with alternating signs. 

The drawback of both schemes is the violation of Adler-Bardeen theorem for  the non-renormalization of the axial anomaly. This must be fixed by an extra finite renormalization constant $Z_{qq}^5$, derived from an external condition, see detailed discussion in \cite{Larin:1993tq,Matiounine:1998re,Ravindran:2003gi}. The NNLO calculation of polarized deep-inelastic-scattering and Drell-Yan process in refs.\cite{Matiounine:1998re,Ravindran:2003gi} made in (HVBM) have shown that the finite renormalization is required only for the the quark-to-quark part (both singlet and non-singlet cases). The same finite renormalization constant can be used for Larin scheme up to $\epsilon$-singular terms at NNLO \cite{Moch:2014sna}. However, it seems that for higher order terms (in $\epsilon$ or in the coupling constant) the constant should be modified \cite{Moch:2014sna}. 

Needless to say, that Larin scheme is far more convenient then HVBM, because it does not violate Lorentz invariance. However, Larin scheme, as it is originally formulated and used in the modern applications \cite{Moch:2014sna}, is inapplicable for TMD calculations. The point is that it does violate the definition of the leading dynamical twist Eq.~(\ref{gG=0}). Indeed, in the Larin scheme we have
\begin{eqnarray}
\gamma^+\Gamma=\gamma^+\(\gamma^+\gamma^5\)_{\text{Larin}}=\frac{i}{3!}\epsilon^{+\nu\alpha\beta}\gamma^+\gamma_{\nu}\gamma_{\alpha}\gamma_{\beta}\neq 0,
\end{eqnarray}
because there is a contribution when all indices $\{\nu\alpha\beta\}$ are transverse. Note, that in HVBM scheme there is not such problem, since in the 4-dimensional $\epsilon^{+\nu\alpha\beta}$, one of the indices is necessarily "$-$". To ensure the existence of Eq.~(\ref{gG=0}) we perform a light modification of Larin scheme, and call it \textit{Larin$^+$} scheme. We define
\begin{eqnarray}\label{Larin+}
(\gamma^+\gamma^5)_{\text{Larin}^+}=\frac{i\epsilon^{+-\alpha\beta}}{2!}\gamma^+\gamma_\alpha\gamma_\beta=\frac{i\epsilon_T^{\alpha\beta}}{2!}\gamma^+\gamma_\alpha\gamma_\beta.
\end{eqnarray}
The $\epsilon_T$-tensor is $d$-dimensional, and for calculations it should be supplemented by the relation
\begin{eqnarray}
\epsilon_T^{\alpha_1\beta_1}\epsilon_T^{\alpha_2\beta_2}=-g_T^{\alpha_1\alpha_2}g_T^{\beta_1\beta_2}+g_T^{\alpha_1\beta_2}g_T^{\beta_1\alpha_2}.
\end{eqnarray}
In the case the $\epsilon$-tensor is 4-dimensional, the definition Eq.~(\ref{Larin+}) coincides with HVBM. The normalization factors presented in the Eq.~(\ref{hel:proj}) are
\begin{eqnarray}
\mathcal{N}_{\text{sch.}}&=&\left\{\begin{array}{cc}
\Ds{1}& \text{HVBM},
\\
\Ds{(1-\epsilon)^{-1}(1-2\epsilon)^{-1}} & \text{Larin}^+.
\end{array}\right.
\end{eqnarray}

The NLO calculation is straightforward and parallel to unpolarized calculation, which is presented in details in \cite{Echevarria:2016scs}. We write  the matching onto integrated distribution as
\begin{align}
R\Phi_{q}^{[\gamma^+\gamma_5]}&=\Delta C_{q\leftarrow q}\otimes \phi_{q}^{[\gamma^+\gamma_5]}+
\Delta C_{q\leftarrow g}\otimes \phi_{g}^{[\epsilon_T^{}]}\nn\\
R\Phi_{g}^{[\epsilon_T^{}]}&=\Delta C_{g\leftarrow q}\otimes \phi_{q}^{[\gamma^+\gamma_5]}+
\Delta C_{g\leftarrow g}\otimes \phi_{g}^{[\epsilon_T^{}]}\
\end{align}
where
%\begin{widetext}
\begin{align}
\nn
 \Delta C_{q\ot q}&=\delta(\bar x)+a_s C_F
\Bigg\{2\pmb B^{\epsilon}\Gamma(-\epsilon)\Big[\frac{2}{(1-x)_+}-2
\\&\nn
+\bar x(1+\epsilon)\mathcal{H}_{\text{sch.}}+\delta(\bar x)\(
\mathbf{L}_{\sqrt{\zeta}}-\psi(-\epsilon)-\gamma_E\)\Big]\Bigg\}_{\epsilon\text{-finite}},
\\
\nn
\Delta C_{q\ot g}&=a_s C_F
\Bigg\{2\pmb B^{\epsilon}\Gamma(-\epsilon)\Big[x-\bar x \mathcal{H}_{\text{sch.}}\Big]\Bigg\}_{\epsilon\text{-finite}},
\end{align}
\begin{align}
%\\
 \nn
\Delta C_{g\ot q}&=a_s C_F
\Bigg\{2\pmb B^{\epsilon}\Gamma(-\epsilon)\Big[1+\bar x \mathcal{H}_{\text{sch.}}\Big]\Bigg\}_{\epsilon\text{-finite}},
\\\label{hel:gg}
 \Delta C_{g\ot g}&=\delta(\bar x)+a_s C_A
\Bigg\{2\pmb B^{\epsilon}\Gamma(-\epsilon)\frac{1}{x}\Big[\frac{2}{(1-x)_+}-2
\\\nn&
-2x^2+2x\bar x\mathcal{H}_{\text{sch.}}+
\delta(\bar x)\(
\mathbf{L}_{\sqrt{\zeta}}-\psi(-\epsilon)-\gamma_E\)\Big]\Bigg\}_{\epsilon\text{-finite}},
\end{align}
%\end{widetext}
where the subscript "$\epsilon$-finite" implies the removal of $\epsilon$-singular terms, as discussed in the end of sec.\ref{sec:OPE}. The coefficient $\mathcal{H}_{\text{sch.}}$ accumulates the difference between schemes,
\begin{eqnarray}
\mathcal{H}_{\text{sch.}}&=&\left\{\begin{array}{cc}
\Ds{1+2\epsilon}& \text{HVBM},
\\
\Ds{\frac{1+\epsilon}{1-\epsilon}} & \text{Larin}^+.
\end{array}\right.
\end{eqnarray}
One can see that the expressions within HVBM and Larin$^+$  schemes coincide up to $\epsilon$-suppressed parts.

In the regime of large-$q_T$, the TMD factorization reproduces the collinear factorization. Therefore, it is natural to normalize the helicity TMDPDF such that at large-$q_T$ it reproduces the cross-section for polarized Drell-Yan, which in turn is normalized onto cross-section of unpolarized Drell-Yan process \cite{Ravindran:2003gi}. The TMD equivalent of this statement is the requirement of equality between helicity and unpolarized matching coefficients
\begin{eqnarray}
\Big[Z_{qq}^5(\vec b)\otimes  \Delta C_{q\ot q}(\vec b)\Big](x)=C_{q\ot q}(x,\vec b).
\end{eqnarray}
The constant $Z_{qq}^5$ is universal, in the sense that it is independent on the rapidity regularization scheme. We find the following finite renormalization constant for the TMD matching
\begin{eqnarray}\label{Zqq5}
Z_{qq}^5=\delta(\bar x )+2  a_sC_F\pmb B^\epsilon \Gamma(-\epsilon)\(1-\epsilon-(1+\epsilon)\mathcal{H}_{\text{sch.}}\)\bar x.
\end{eqnarray}
Note, that HVBM version of $Z_{qq}^5$ coincides with the NLO part of the one presented in \cite{Ravindran:2003gi} up to logarithmic terms (which are dependent on the kinematics of process) .  

Concluding the section we present the expressions for the helicity TMD distribution in the regime of small-$b$
\begin{align}\nn
g_{1L}(x,\vec b)&=[\Delta C_{q\ot q}(\vec b)\otimes \Delta f_q](x)+[\Delta C_{q\ot g}(\vec b)\otimes \Delta f_g](x)+\mathcal{O}(\vec b^2),
\\
g^g_{1L}(x,\vec b)&=[\Delta C_{g\ot q}(\vec b)\otimes \Delta f_q](x)+[\Delta C_{g\ot g}(\vec b)\otimes \Delta f_g](x)+\mathcal{O}(\vec b^2),
\end{align}
where the matching coefficients are taken in the limit $\epsilon\rightarrow 0$,
\begin{align}
\Delta C_{q\ot q}&\equiv C_{q\ot q}=\delta(\bar x)+a_s C_F\Big(
-2\mathbf{L}_{\mu}\Delta p_{qq}+2\bar x\nn
\\&\nn +\delta(\bar x)\(-\mathbf{L}_{\mu}^2+2\mathbf{L}_{\mu}\mathbf{l}_\zeta-\zeta_2\)\Big)+\mathcal{O}(a_s^2),
\\
\nn \Delta C_{q\ot g}&=a_s T_F\(-2\mathbf{L}_{\mu} \Delta p_{qg}+4\bar x\)+\mathcal{O}(a_s^2),
\\
\nn \Delta C_{g\ot q}&=a_s C_F\(-2\mathbf{L}_{\mu} \Delta p_{gq}-4\bar x\)+\mathcal{O}(a_s^2),
\\
\nn \Delta C_{g \ot g }&=\delta(\bar x)+a_s C_A\Big(
-2\mathbf{L}_{\mu}\Delta p_{gg}-8\bar x
\\ 
& +\delta(\bar x)\(-\mathbf{L}_{\mu}^2+2\mathbf{L}_{\mu}\mathbf{l}_\zeta-\zeta_2\)\Big)+\mathcal{O}(a_s^2),
\end{align}
with $\mathbf{l}_\zeta=\ln\mu^2/\zeta$. 
The functions $\Delta p$ are the combination of helicity evolution kernel (which can be found e.g. in~\cite{Moch:2014sna}) and the TMD anomalous dimension. They are
\begin{align}\label{app:hel_qq}
\Delta p_{qq}(x)&=\frac{2}{(1-x)_+}-1-x
,\nn
\\
\Delta p_{qg}(x)&=2x-1,\nn
\quad
\Delta p_{gq}(x)=2-x,
\\ %\label{app:hel_gg}
\Delta p_{gg}(x)&=\frac{2}{(1-x)_+}+2-4x.
\end{align}
The coefficients $\Delta C_{q\ot q}$ and $\Delta C_{q\ot g}$ have been evaluated in \cite{Bacchetta:2013pqa}. Our expressions agree with ones presented in \cite{Bacchetta:2013pqa} apart of $\zeta_2$ term in $\Delta C_{q\ot q}$. This disagreement is the result of different renormalization schemes. We use the conventional $\overline{\text{MS}}$ scheme with $e^{\epsilon\gamma_E}$ factor, while $\text{MS}$-scheme of \cite{Bacchetta:2013pqa} is defined with $\Gamma^{-1}(1+\epsilon)$ factor. The coefficients $\Delta C_{g\ot q}$ and $\Delta C_{g\ot g}$ have been evaluated in \cite{Echevarria:2015uaa}. Our expressions agree with expressions presented in the erratum of Ref.\cite{Echevarria:2015uaa}.

\section{Transversity and pretzelosity distributions}
\label{sec:trans_and_pretz}
The spinor structure for the transversity TMD operator is usually addressed as $\Gamma=i\gamma_5\sigma^{+\mu}=\epsilon_T^{\mu\nu} \sigma^{+\nu}/2,$ where to obtain the last equality we used that index $\mu$ is transverse. This definition is scheme dependent just as the helicity case. However, since there is no mixture with the gluons at the leading twist, the common practice is to eliminate the $\gamma^5$ or $\epsilon_T$ from the definition of operator. Thus we consider $\Gamma=\sigma^{+\mu}$. The small-$b$ expansion takes the form
%\begin{widetext}
\begin{align}\label{trans->trans}
R\Phi_{q}^{[\sigma^{+\mu}]}&=\Big\{\delta(\bar x)g_T^{\mu\nu}+2a_s C_F\pmb B^\epsilon\Gamma(-\epsilon)\Big[
\\&\nn
g_T^{\mu\nu}\(\frac{2}{(1-x)_+}-2+\delta(\bar x)(\mathbf{L}_{\sqrt{\zeta}}-\psi(-\epsilon)-\gamma_E)\)
\\&\nn \qquad\qquad\qquad\qquad-
2\epsilon^2\bar x \frac{ b^\mu  b^\nu}{\vec b^2}\Big]\Big\}\otimes \phi^{[\sigma^{+\nu}]}_{q}.
\end{align}
Comparing this expression with the parameterization Eq.~(\ref{TMD_Q_dec}) we observe that both the transversity distribution and the pretzelosity distributions have the leading twist-2 matching on the integrated transversity PDF.

The transversity and pretzelosity distribution matching coefficients, respectively $ \delta C_{q\ot q}$ and $\delta^\perp C_{q\ot q}$,
are defined as 
\begin{align}\label{trans_OPE}
R\Phi_{q}^{[\sigma^{+\mu}]}&= g^{\mu\nu}_T \delta C_{q\ot q}\otimes \phi_{q}^{[\sigma^{+\nu}]}
\\\nn&
+\(\frac{b^\mu b^\nu}{\vec b^2}+\frac{g_T^{\mu\nu}}{2(1-\epsilon)}\)\delta^\perp C_{q\ot q}\otimes \phi_{q}^{[\sigma^{+\nu}]},
\end{align}
where the factor $(1-\epsilon)$ in the pretzelosity vector structure is necessary to support its tracelessness in  dimensional regularization.

Comparing expressions (\ref{trans_OPE}) with (\ref{trans->trans}) we obtain
\begin{align}
 \delta C_{q\ot q}&=\delta(\bar x)+a_s C_F
\Bigg\{2\pmb B^{\epsilon}\Gamma(-\epsilon)\Big[\frac{2}{(1-x)_+}-2
\\\nn &+\bar x \frac{\epsilon^2}{1-\epsilon}+\delta(\bar x)\(
\mathbf{L}_{\sqrt{\zeta}}-\psi(-\epsilon)-\gamma_E\)\Big]\Bigg\}_{\epsilon\text{-finite}}.
\end{align}
It results to the following small-$b$ expression for the transversity TMD PDF
\begin{eqnarray}
h_{1}(x,\vec b)&=\Big[\delta C_{q\ot q}(\vec b)\otimes \delta f_q\Big](x)+\mathcal{O}(\vec b^2),
\end{eqnarray}
with the matching coefficient
\begin{align}
 \delta C_{q\ot q}&=\delta(\bar x)+a_s C_F\Big(
-2\mathbf{L}_{\mu}\delta p_{qq}
 \\\nn&
+\delta(\bar x)\(-\mathbf{L}_{\mu}^2+2\mathbf{L}_{\mu}\mathbf{l}_\zeta-\zeta_2\)\Big)+\mathcal{O}(a_s^2).
\end{align}
The $\delta p_{qq}$ is the combination of the transversity evolution kernel (see e.g.\cite{Vogelsang:1997ak,Mikhailov:2008my}) and  TMD anomalous dimension. It is
\begin{eqnarray}\label{app:trans}
\delta p_{qq}(x)=\frac{2}{(1-x)_+}-2.
\end{eqnarray}
This expression coincides with the one calculated in \cite{Bacchetta:2013pqa} up to $\zeta_2$ term (which is absent in \cite{Bacchetta:2013pqa} due to the usage of a different form of $\overline{\text{MS}}$-scheme).

The matching coefficient of the pretzelosity distribution at finite $\epsilon$ is
\begin{eqnarray}
\delta^\perp C_{q\ot q}=-4a_s C_F\pmb B^\epsilon \Gamma(-\epsilon)\bar x \epsilon^2.
\end{eqnarray}
Here, we can appreciate the consistent and natural counting of the normalization of the pretzelosity provided  by Eq.~(\ref{TMD_Q_dec}). We also observe that at this order of perturbation theory the matching coefficient is proportional to $\epsilon$, i.e. zero. 
Nonetheless, the $\epsilon$-suppressed part will reveal at NNLO, and provide a non-zero contribution. Therefore, we conclude
\begin{align}\label{pretz}\nn
h_{1T}^\perp(x,\vec b)&=\Big[\delta^\perp C_{q\ot q}(\vec b)\otimes \delta f_q\Big](x)+\mathcal{O}(\vec b^2)=
\\
& \Big[ \(0+\mathcal{O}(a_s^2)\)\otimes \delta f_q\Big](x)+\mathcal{O}(\vec b^2).
\end{align}
This result coincides with the estimation made in \cite{Bacchetta:2008xw}. According to Eq.~(\ref{pretz}), the pretzelosity distribution is suppressed numerically. This observation is indeed supported by the measurements of $\sin(3\phi_h-\phi_S)$-asymmetries by HERMES and COMPASS, see e.g.\cite{Lefky:2014eia} and references within. We also mention that it is not possible to obtain the small-$b$ matching at the helicity distribution. The helicity distribution as a part of pretzelosity distribution is suggested by various model calculations (see \cite{Avakian:2008dz} and references within).

%%%%%%%%%%%%%%%%%%%%%
\section{Linearly polarized gluon}
%%%%%%%%%%%%%%%%%%%%%

The linearly polarized gluon distribution at small-$b$ matches the unpolarized gluon distribution. The matching of the gluon TMD operator to the unpolarized distribution has the form
\begin{align}
R\Phi_{g}^{\mu\nu}=\(\frac{b^\mu b^\nu}{\vec b^2}+\frac{g_T^{\mu\nu}}{2(1-\epsilon)}\)\Big(
&\delta^L C_{g\ot g}\otimes \phi_g^{[g_T]}
\\\nn&+\delta^L C_{g\ot q}\otimes \phi_q^{[\gamma^+]}\Big)+...~,
\end{align}
where dots represent terms proportional to $g_T^{\mu\nu}$ and $\epsilon_T^{\mu\nu}$, i.e. the parts which contribute to the matching of unpolarized and helicity distributions. 

The coefficients $\delta^L C$ are\footnote{We thank M. Diehl for pointing out a sign typo in eq.~(\ref{eq:dc1}, \ref{eq:dc2}, \ref{eq:dc3}) in the previous version of the paper.}
\begin{eqnarray}
\label{eq:dc1}
\delta^L C_{g\ot g}&=&\Big(+4a_s C_A \pmb B^\epsilon \Gamma(-\epsilon)\frac{\bar x}{x}\epsilon\Big)_{\epsilon\text{-finite}}.
\nn
\\
\delta^LC_{g\ot q}&=&\Big(+4a_s C_F \pmb B^\epsilon \Gamma(-\epsilon)\frac{\bar x}{x}\epsilon\Big)_{\epsilon\text{-finite}}.
\end{eqnarray}
Note that there is not rapidity nor renormalization group evolution, which appears at the next perturbative order. 

Finally, we obtain following small-$b$ expression for the linearly polarized gluon TMDPDF
\begin{align}
h_1^{\perp g}(x,\vec b) &=
\\ \nn &
[\delta^LC_{g\ot q}(\vec b)\otimes f_q](x)+[\delta^L C_{g\ot g}(\vec b)\otimes f_g](x)+\mathcal{O}(\vec b^2),
\end{align}
where
\begin{eqnarray}
\label{eq:dc2}
\delta^LC_{g\ot g}&=&-4a_s C_A \frac{\bar x}{x}+\mathcal{O}(a_s^2)
,
\\
\label{eq:dc3}
\delta^LC_{g\ot q}&=&-4a_s C_F \frac{\bar x}{x}+\mathcal{O}(a_s^2).
\end{eqnarray}
%These matching coefficients agree with  \cite{Echevarria:2015uaa}.

%%%%%%%%%%%%%%%%%%%%%%%
\section{Conclusions}
 \label{sec:Conclusions}
%%%%%%%%%%%%%%%%%%

In this letter, we have provided complete discussion on the matching of transverse momentum dependent (TMD) distributions to the twist-2 integrated distributions in the regime of small-$b$ (or equivalently, large-$q_T$). To perform the matching we have evaluated the operator product expansion (OPE) of a generic TMD operator near the light-cone. 

As a practical outcome, we derive the complete set of NLO TMD matching coefficients of the twist-2 parton distributions evaluated uniformly at finite $\epsilon$. The TMD distributions that have non-zero matching are helicity ($g_{1L}$, $g_{1L}^g$), transvesity ($h_1$), pretzelosity ($h_{1T}^\perp$) and linearly polarized gluon ($h_1^{\perp g}$) distributions (we do not include the unpolarized TMD distribution in the consideration because it has been considered in many articles. The evaluation performed using  the same regularization as this paper can be found in \cite{Echevarria:2016scs}). The most part of the coefficient functions have been evaluated separately for quarks and gluons by different groups \cite{Bacchetta:2013pqa,Echevarria:2015uaa}. We agree with their evaluations (taking into account that in ref.\cite{Bacchetta:2013pqa}, different renormalization scheme has been used).

The evaluation of OPE for a generic TMD operator reveals the condition which should be satisfied in order match the rapidity divergences of a TMD operator and the leading order TMD soft factor Eq.~(\ref{eq:softf}). The conditions presented in Eqs.~(\ref{gG=0},\ref{Lambda+=0}) restrict the Lorentz structure of the TMD operators.
 The TMD distributions whose operator meet these conditions, are known as TMD distributions of leading dynamical twist. 
In this way, we demonstrate that the next-to-leading-dynamical-twist contributions to the TMD factorization theorem (i.e. the power suppressed contributions to the TMD cross-section) necessarily have a different structure of rapidity divergences.

We also provide  discussion on the schemes of $\gamma_5$ and $\epsilon_T$-definition in the dimensional regularization, which has been skipped by the previous authors. We have shown that the definition of $\gamma_5$ suggested by the popular Larin scheme \cite{Larin:1993tq} does not support the condition of the leading dynamical twist, and thus, it is inapplicable in TMD calculations. We suggest an updated version of Larin scheme (Larin$^+$ scheme Eq.~(\ref{Larin+})), which supports the condition and has simpler properties than the traditional one. Our calculation has been performed in Larin$^+$ and HVBM \cite{tHooft:1972tcz,Breitenlohner:1977hr} schemes. At NLO the difference between schemes arises only in the $\epsilon$-suppressed terms.
 We argue about  the normalization of the distributions and derive the finite renormalization constant (\ref{Zqq5}) for TMD helicity distributions in both schemes.

The evaluation of the matching has been performed at finite-$\epsilon$. The $\epsilon$-suppressed terms, although do not contribute directly to NLO, contribute to higher perturbative orders (see e.g. discussion in \cite{Echevarria:2016scs}). The pretzelosity distribution (considered here for the first time) has $\epsilon$-suppressed matching coefficient, which indicates that it has non-zero matching coefficient to transversity distribution at NNLO Eq.~(\ref{pretz}). This offers a natural explanation of the smallness of this distribution in phenomenological analyses~\cite{Lefky:2014eia}. The complete $\epsilon$-dependent expressions and the general analyses performed in this work open the path to the NNLO evaluation of polarized TMD distributions.

%%%%%%%%%%%%%%%%%%%%%%%%%%%%%%%%%
%%%%%%%%%%%%%%%%%%%%%%%%%%%%%%%%%
%%%%%%%%%%%%%%%%%%%%%%%%%%%%%%%%%
\section*{Acknowledgements}
%%%%%%%%%%%%%%%%%%%%%%%%%%%%%%%%%
We acknowledge some very profitable discussions with V. Braun, M. G. A. Buffing, M. Diehl, M. G. Echevarria, T. Kasemets as well as, S.Moch for discussions on the definition of $\gamma_5$. D.G.R. and I.S. are supported by the Spanish MECD grant FPA2014-53375-C2-2-P and  FPA2016-75654-C2-2-P.
%%%%%%%%%%%%%%%%%%%%%%%%%%%%%%%%%
%%%%%%%%%%%%%%%%%%%%%%%%%%%%%%%%%
%%%%%%%%%%%%%%%%%%%%%%%%%%%%%%%%%
\bibliographystyle{apsrev4-1}
\bibliography{Refbib}

\begin{thebibliography}{32}%
\makeatletter
\providecommand \@ifxundefined [1]{%
 \@ifx{#1\undefined}
}%
\providecommand \@ifnum [1]{%
 \ifnum #1\expandafter \@firstoftwo
 \else \expandafter \@secondoftwo
 \fi
}%
\providecommand \@ifx [1]{%
 \ifx #1\expandafter \@firstoftwo
 \else \expandafter \@secondoftwo
 \fi
}%
\providecommand \natexlab [1]{#1}%
\providecommand \enquote  [1]{``#1''}%
\providecommand \bibnamefont  [1]{#1}%
\providecommand \bibfnamefont [1]{#1}%
\providecommand \citenamefont [1]{#1}%
\providecommand \href@noop [0]{\@secondoftwo}%
\providecommand \href [0]{\begingroup \@sanitize@url \@href}%
\providecommand \@href[1]{\@@startlink{#1}\@@href}%
\providecommand \@@href[1]{\endgroup#1\@@endlink}%
\providecommand \@sanitize@url [0]{\catcode `\\12\catcode `\$12\catcode
  `\&12\catcode `\#12\catcode `\^12\catcode `\_12\catcode `\%12\relax}%
\providecommand \@@startlink[1]{}%
\providecommand \@@endlink[0]{}%
\providecommand \url  [0]{\begingroup\@sanitize@url \@url }%
\providecommand \@url [1]{\endgroup\@href {#1}{\urlprefix }}%
\providecommand \urlprefix  [0]{URL }%
\providecommand \Eprint [0]{\href }%
\providecommand \doibase [0]{http://dx.doi.org/}%
\providecommand \selectlanguage [0]{\@gobble}%
\providecommand \bibinfo  [0]{\@secondoftwo}%
\providecommand \bibfield  [0]{\@secondoftwo}%
\providecommand \translation [1]{[#1]}%
\providecommand \BibitemOpen [0]{}%
\providecommand \bibitemStop [0]{}%
\providecommand \bibitemNoStop [0]{.\EOS\space}%
\providecommand \EOS [0]{\spacefactor3000\relax}%
\providecommand \BibitemShut  [1]{\csname bibitem#1\endcsname}%
\let\auto@bib@innerbib\@empty
%</preamble>
\bibitem [{\citenamefont {Echevarria}\ \emph {et~al.}(2012)\citenamefont
  {Echevarria}, \citenamefont {Idilbi},\ and\ \citenamefont
  {Scimemi}}]{GarciaEchevarria:2011rb}%
  \BibitemOpen
  \bibfield  {author} {\bibinfo {author} {\bibfnamefont {M.~G.}\ \bibnamefont
  {Echevarria}}, \bibinfo {author} {\bibfnamefont {A.}~\bibnamefont {Idilbi}},
  \ and\ \bibinfo {author} {\bibfnamefont {I.}~\bibnamefont {Scimemi}},\ }\href
  {\doibase 10.1007/JHEP07(2012)002} {\bibfield  {journal} {\bibinfo  {journal}
  {JHEP}\ }\textbf {\bibinfo {volume} {07}},\ \bibinfo {pages} {002} (\bibinfo
  {year} {2012})},\ \Eprint {http://arxiv.org/abs/1111.4996} {arXiv:1111.4996
  [hep-ph]} \BibitemShut {NoStop}%
%%CITATION = ARXIV:1111.4996;%%
\bibitem [{\citenamefont {Collins}(2013)}]{Collins:2011zzd}%
  \BibitemOpen
  \bibfield  {author} {\bibinfo {author} {\bibfnamefont {J.}~\bibnamefont
  {Collins}},\ }\href {http://www.cambridge.org/de/knowledge/isbn/item5756723}
  {\emph {\bibinfo {title} {{Foundations of perturbative QCD}}}}\ (\bibinfo
  {publisher} {Cambridge University Press},\ \bibinfo {year}
  {2013})\BibitemShut {NoStop}%
%%CITATION = INSPIRE-922696;%%
\bibitem [{\citenamefont {Chiu}\ \emph {et~al.}(2012)\citenamefont {Chiu},
  \citenamefont {Jain}, \citenamefont {Neill},\ and\ \citenamefont
  {Rothstein}}]{Chiu:2012ir}%
  \BibitemOpen
  \bibfield  {author} {\bibinfo {author} {\bibfnamefont {J.-Y.}\ \bibnamefont
  {Chiu}}, \bibinfo {author} {\bibfnamefont {A.}~\bibnamefont {Jain}}, \bibinfo
  {author} {\bibfnamefont {D.}~\bibnamefont {Neill}}, \ and\ \bibinfo {author}
  {\bibfnamefont {I.~Z.}\ \bibnamefont {Rothstein}},\ }\href {\doibase
  10.1007/JHEP05(2012)084} {\bibfield  {journal} {\bibinfo  {journal} {JHEP}\
  }\textbf {\bibinfo {volume} {1205}},\ \bibinfo {pages} {084} (\bibinfo {year}
  {2012})},\ \Eprint {http://arxiv.org/abs/1202.0814} {arXiv:1202.0814
  [hep-ph]} \BibitemShut {NoStop}%
%%CITATION = ARXIV:1202.0814;%%
\bibitem [{\citenamefont {Echevarria}\ \emph {et~al.}(2014)\citenamefont
  {Echevarria}, \citenamefont {Idilbi},\ and\ \citenamefont
  {Scimemi}}]{Echevarria:2014rua}%
  \BibitemOpen
  \bibfield  {author} {\bibinfo {author} {\bibfnamefont {M.~G.}\ \bibnamefont
  {Echevarria}}, \bibinfo {author} {\bibfnamefont {A.}~\bibnamefont {Idilbi}},
  \ and\ \bibinfo {author} {\bibfnamefont {I.}~\bibnamefont {Scimemi}},\ }\href
  {\doibase 10.1103/PhysRevD.90.014003} {\bibfield  {journal} {\bibinfo
  {journal} {Phys. Rev.}\ }\textbf {\bibinfo {volume} {D90}},\ \bibinfo {pages}
  {014003} (\bibinfo {year} {2014})},\ \Eprint {http://arxiv.org/abs/1402.0869}
  {arXiv:1402.0869 [hep-ph]} \BibitemShut {NoStop}%
%%CITATION = ARXIV:1402.0869;%%
\bibitem [{\citenamefont {Becher}\ and\ \citenamefont
  {Neubert}(2011)}]{Becher:2010tm}%
  \BibitemOpen
  \bibfield  {author} {\bibinfo {author} {\bibfnamefont {T.}~\bibnamefont
  {Becher}}\ and\ \bibinfo {author} {\bibfnamefont {M.}~\bibnamefont
  {Neubert}},\ }\href {\doibase 10.1140/epjc/s10052-011-1665-7} {\bibfield
  {journal} {\bibinfo  {journal} {Eur. Phys. J.}\ }\textbf {\bibinfo {volume}
  {C71}},\ \bibinfo {pages} {1665} (\bibinfo {year} {2011})},\ \Eprint
  {http://arxiv.org/abs/1007.4005} {arXiv:1007.4005 [hep-ph]} \BibitemShut
  {NoStop}%
%%CITATION = ARXIV:1007.4005;%%
\bibitem [{\citenamefont {Aybat}\ and\ \citenamefont
  {Rogers}(2011)}]{Aybat:2011zv}%
  \BibitemOpen
  \bibfield  {author} {\bibinfo {author} {\bibfnamefont {S.~M.}\ \bibnamefont
  {Aybat}}\ and\ \bibinfo {author} {\bibfnamefont {T.~C.}\ \bibnamefont
  {Rogers}},\ }\href {\doibase 10.1103/PhysRevD.83.114042} {\bibfield
  {journal} {\bibinfo  {journal} {Phys. Rev.}\ }\textbf {\bibinfo {volume}
  {D83}},\ \bibinfo {pages} {114042} (\bibinfo {year} {2011})},\ \Eprint
  {http://arxiv.org/abs/1101.5057} {arXiv:1101.5057 [hep-ph]} \BibitemShut
  {NoStop}%
%%CITATION = ARXIV:1101.5057;%%
\bibitem [{\citenamefont {Cherednikov}\ and\ \citenamefont
  {Stefanis}(2008{\natexlab{a}})}]{Cherednikov:2008ua}%
  \BibitemOpen
  \bibfield  {author} {\bibinfo {author} {\bibfnamefont {I.~O.}\ \bibnamefont
  {Cherednikov}}\ and\ \bibinfo {author} {\bibfnamefont {N.~G.}\ \bibnamefont
  {Stefanis}},\ }\href {\doibase 10.1016/j.nuclphysb.2008.05.011} {\bibfield
  {journal} {\bibinfo  {journal} {Nucl. Phys.}\ }\textbf {\bibinfo {volume}
  {B802}},\ \bibinfo {pages} {146} (\bibinfo {year} {2008}{\natexlab{a}})},\
  \Eprint {http://arxiv.org/abs/0802.2821} {arXiv:0802.2821 [hep-ph]}
  \BibitemShut {NoStop}%
%%CITATION = ARXIV:0802.2821;%%
\bibitem [{\citenamefont {Cherednikov}\ and\ \citenamefont
  {Stefanis}(2008{\natexlab{b}})}]{Cherednikov:2007tw}%
  \BibitemOpen
  \bibfield  {author} {\bibinfo {author} {\bibfnamefont {I.~O.}\ \bibnamefont
  {Cherednikov}}\ and\ \bibinfo {author} {\bibfnamefont {N.~G.}\ \bibnamefont
  {Stefanis}},\ }\href {\doibase 10.1103/PhysRevD.77.094001} {\bibfield
  {journal} {\bibinfo  {journal} {Phys. Rev.}\ }\textbf {\bibinfo {volume}
  {D77}},\ \bibinfo {pages} {094001} (\bibinfo {year} {2008}{\natexlab{b}})},\
  \Eprint {http://arxiv.org/abs/0710.1955} {arXiv:0710.1955 [hep-ph]}
  \BibitemShut {NoStop}%
%%CITATION = ARXIV:0710.1955;%%
\bibitem [{\citenamefont {Echevarria}\ \emph
  {et~al.}(2016{\natexlab{a}})\citenamefont {Echevarria}, \citenamefont
  {Scimemi},\ and\ \citenamefont {Vladimirov}}]{Echevarria:2015usa}%
  \BibitemOpen
  \bibfield  {author} {\bibinfo {author} {\bibfnamefont {M.~G.}\ \bibnamefont
  {Echevarria}}, \bibinfo {author} {\bibfnamefont {I.}~\bibnamefont {Scimemi}},
  \ and\ \bibinfo {author} {\bibfnamefont {A.}~\bibnamefont {Vladimirov}},\
  }\href {\doibase 10.1103/PhysRevD.93.011502} {\bibfield  {journal} {\bibinfo
  {journal} {Phys. Rev.}\ }\textbf {\bibinfo {volume} {D93}},\ \bibinfo {pages}
  {011502} (\bibinfo {year} {2016}{\natexlab{a}})},\ \Eprint
  {http://arxiv.org/abs/1509.06392} {arXiv:1509.06392 [hep-ph]} \BibitemShut
  {NoStop}%
%%CITATION = ARXIV:1509.06392;%%
\bibitem [{\citenamefont {Echevarria}\ \emph
  {et~al.}(2016{\natexlab{b}})\citenamefont {Echevarria}, \citenamefont
  {Scimemi},\ and\ \citenamefont {Vladimirov}}]{Echevarria:2016scs}%
  \BibitemOpen
  \bibfield  {author} {\bibinfo {author} {\bibfnamefont {M.~G.}\ \bibnamefont
  {Echevarria}}, \bibinfo {author} {\bibfnamefont {I.}~\bibnamefont {Scimemi}},
  \ and\ \bibinfo {author} {\bibfnamefont {A.}~\bibnamefont {Vladimirov}},\
  }\href {\doibase 10.1007/JHEP09(2016)004} {\  (\bibinfo {year}
  {2016}{\natexlab{b}}),\ 10.1007/JHEP09(2016)004},\ \Eprint
  {http://arxiv.org/abs/1604.07869} {arXiv:1604.07869 [hep-ph]} \BibitemShut
  {NoStop}%
%%CITATION = ARXIV:1604.07869;%%
\bibitem [{\citenamefont {Gehrmann}\ \emph {et~al.}(2014)\citenamefont
  {Gehrmann}, \citenamefont {Luebbert},\ and\ \citenamefont
  {Yang}}]{Gehrmann:2014yya}%
  \BibitemOpen
  \bibfield  {author} {\bibinfo {author} {\bibfnamefont {T.}~\bibnamefont
  {Gehrmann}}, \bibinfo {author} {\bibfnamefont {T.}~\bibnamefont {Luebbert}},
  \ and\ \bibinfo {author} {\bibfnamefont {L.~L.}\ \bibnamefont {Yang}},\
  }\href {\doibase 10.1007/JHEP06(2014)155} {\bibfield  {journal} {\bibinfo
  {journal} {JHEP}\ }\textbf {\bibinfo {volume} {06}},\ \bibinfo {pages} {155}
  (\bibinfo {year} {2014})},\ \Eprint {http://arxiv.org/abs/1403.6451}
  {arXiv:1403.6451 [hep-ph]} \BibitemShut {NoStop}%
%%CITATION = ARXIV:1403.6451;%%
\bibitem [{\citenamefont {Gehrmann}\ \emph {et~al.}(2012)\citenamefont
  {Gehrmann}, \citenamefont {Lubbert},\ and\ \citenamefont
  {Yang}}]{Gehrmann:2012ze}%
  \BibitemOpen
  \bibfield  {author} {\bibinfo {author} {\bibfnamefont {T.}~\bibnamefont
  {Gehrmann}}, \bibinfo {author} {\bibfnamefont {T.}~\bibnamefont {Lubbert}}, \
  and\ \bibinfo {author} {\bibfnamefont {L.~L.}\ \bibnamefont {Yang}},\ }\href
  {\doibase 10.1103/PhysRevLett.109.242003} {\bibfield  {journal} {\bibinfo
  {journal} {Phys. Rev. Lett.}\ }\textbf {\bibinfo {volume} {109}},\ \bibinfo
  {pages} {242003} (\bibinfo {year} {2012})},\ \Eprint
  {http://arxiv.org/abs/1209.0682} {arXiv:1209.0682 [hep-ph]} \BibitemShut
  {NoStop}%
%%CITATION = ARXIV:1209.0682;%%
\bibitem [{\citenamefont {Bacchetta}\ and\ \citenamefont
  {Prokudin}(2013)}]{Bacchetta:2013pqa}%
  \BibitemOpen
  \bibfield  {author} {\bibinfo {author} {\bibfnamefont {A.}~\bibnamefont
  {Bacchetta}}\ and\ \bibinfo {author} {\bibfnamefont {A.}~\bibnamefont
  {Prokudin}},\ }\href {\doibase 10.1016/j.nuclphysb.2013.07.013} {\bibfield
  {journal} {\bibinfo  {journal} {Nucl. Phys.}\ }\textbf {\bibinfo {volume}
  {B875}},\ \bibinfo {pages} {536} (\bibinfo {year} {2013})},\ \Eprint
  {http://arxiv.org/abs/1303.2129} {arXiv:1303.2129 [hep-ph]} \BibitemShut
  {NoStop}%
%%CITATION = ARXIV:1303.2129;%%
\bibitem [{\citenamefont {Echevarria}\ \emph {et~al.}(2015)\citenamefont
  {Echevarria}, \citenamefont {Kasemets}, \citenamefont {Mulders},\ and\
  \citenamefont {Pisano}}]{Echevarria:2015uaa}%
  \BibitemOpen
  \bibfield  {author} {\bibinfo {author} {\bibfnamefont {M.~G.}\ \bibnamefont
  {Echevarria}}, \bibinfo {author} {\bibfnamefont {T.}~\bibnamefont
  {Kasemets}}, \bibinfo {author} {\bibfnamefont {P.~J.}\ \bibnamefont
  {Mulders}}, \ and\ \bibinfo {author} {\bibfnamefont {C.}~\bibnamefont
  {Pisano}},\ }\href {\doibase 10.1007/JHEP07(2015)158} {\bibfield  {journal}
  {\bibinfo  {journal} {JHEP}\ }\textbf {\bibinfo {volume} {07}},\ \bibinfo
  {pages} {158} (\bibinfo {year} {2015})},\ \Eprint
  {http://arxiv.org/abs/1502.05354} {arXiv:1502.05354 [hep-ph]} \BibitemShut
  {NoStop}%
%%CITATION = ARXIV:1502.05354;%%
\bibitem [{\citenamefont {Echevarria}\ \emph
  {et~al.}(2016{\natexlab{c}})\citenamefont {Echevarria}, \citenamefont
  {Scimemi},\ and\ \citenamefont {Vladimirov}}]{Echevarria:2015byo}%
  \BibitemOpen
  \bibfield  {author} {\bibinfo {author} {\bibfnamefont {M.~G.}\ \bibnamefont
  {Echevarria}}, \bibinfo {author} {\bibfnamefont {I.}~\bibnamefont {Scimemi}},
  \ and\ \bibinfo {author} {\bibfnamefont {A.}~\bibnamefont {Vladimirov}},\
  }\href {\doibase 10.1103/PhysRevD.93.054004} {\bibfield  {journal} {\bibinfo
  {journal} {Phys. Rev.}\ }\textbf {\bibinfo {volume} {D93}},\ \bibinfo {pages}
  {054004} (\bibinfo {year} {2016}{\natexlab{c}})},\ \Eprint
  {http://arxiv.org/abs/1511.05590} {arXiv:1511.05590 [hep-ph]} \BibitemShut
  {NoStop}%
%%CITATION = ARXIV:1511.05590;%%
\bibitem [{\citenamefont {Vladimirov}(2016)}]{Vladimirov:2016dll}%
  \BibitemOpen
  \bibfield  {author} {\bibinfo {author} {\bibfnamefont {A.~A.}\ \bibnamefont
  {Vladimirov}},\ }\href@noop {} {\  (\bibinfo {year} {2016})},\ \Eprint
  {http://arxiv.org/abs/1610.05791} {arXiv:1610.05791 [hep-ph]} \BibitemShut
  {NoStop}%
%%CITATION = ARXIV:1610.05791;%%
\bibitem [{\citenamefont {Goeke}\ \emph {et~al.}(2005)\citenamefont {Goeke},
  \citenamefont {Metz},\ and\ \citenamefont {Schlegel}}]{Goeke:2005hb}%
  \BibitemOpen
  \bibfield  {author} {\bibinfo {author} {\bibfnamefont {K.}~\bibnamefont
  {Goeke}}, \bibinfo {author} {\bibfnamefont {A.}~\bibnamefont {Metz}}, \ and\
  \bibinfo {author} {\bibfnamefont {M.}~\bibnamefont {Schlegel}},\ }\href
  {\doibase 10.1016/j.physletb.2005.05.037} {\bibfield  {journal} {\bibinfo
  {journal} {Phys. Lett.}\ }\textbf {\bibinfo {volume} {B618}},\ \bibinfo
  {pages} {90} (\bibinfo {year} {2005})},\ \Eprint
  {http://arxiv.org/abs/hep-ph/0504130} {arXiv:hep-ph/0504130 [hep-ph]}
  \BibitemShut {NoStop}%
%%CITATION = HEP-PH/0504130;%%
\bibitem [{\citenamefont {Bacchetta}\ \emph {et~al.}(2008)\citenamefont
  {Bacchetta}, \citenamefont {Boer}, \citenamefont {Diehl},\ and\ \citenamefont
  {Mulders}}]{Bacchetta:2008xw}%
  \BibitemOpen
  \bibfield  {author} {\bibinfo {author} {\bibfnamefont {A.}~\bibnamefont
  {Bacchetta}}, \bibinfo {author} {\bibfnamefont {D.}~\bibnamefont {Boer}},
  \bibinfo {author} {\bibfnamefont {M.}~\bibnamefont {Diehl}}, \ and\ \bibinfo
  {author} {\bibfnamefont {P.~J.}\ \bibnamefont {Mulders}},\ }\href {\doibase
  10.1088/1126-6708/2008/08/023} {\bibfield  {journal} {\bibinfo  {journal}
  {JHEP}\ }\textbf {\bibinfo {volume} {08}},\ \bibinfo {pages} {023} (\bibinfo
  {year} {2008})},\ \Eprint {http://arxiv.org/abs/0803.0227} {arXiv:0803.0227
  [hep-ph]} \BibitemShut {NoStop}%
%%CITATION = ARXIV:0803.0227;%%
\bibitem [{\citenamefont {Mulders}\ and\ \citenamefont
  {Rodrigues}(2001)}]{Mulders:2000sh}%
  \BibitemOpen
  \bibfield  {author} {\bibinfo {author} {\bibfnamefont {P.~J.}\ \bibnamefont
  {Mulders}}\ and\ \bibinfo {author} {\bibfnamefont {J.}~\bibnamefont
  {Rodrigues}},\ }\href {\doibase 10.1103/PhysRevD.63.094021} {\bibfield
  {journal} {\bibinfo  {journal} {Phys. Rev.}\ }\textbf {\bibinfo {volume}
  {D63}},\ \bibinfo {pages} {094021} (\bibinfo {year} {2001})},\ \Eprint
  {http://arxiv.org/abs/hep-ph/0009343} {arXiv:hep-ph/0009343 [hep-ph]}
  \BibitemShut {NoStop}%
%%CITATION = HEP-PH/0009343;%%
\bibitem [{\citenamefont {Boer}\ \emph {et~al.}(2011)\citenamefont {Boer},
  \citenamefont {Gamberg}, \citenamefont {Musch},\ and\ \citenamefont
  {Prokudin}}]{Boer:2011xd}%
  \BibitemOpen
  \bibfield  {author} {\bibinfo {author} {\bibfnamefont {D.}~\bibnamefont
  {Boer}}, \bibinfo {author} {\bibfnamefont {L.}~\bibnamefont {Gamberg}},
  \bibinfo {author} {\bibfnamefont {B.}~\bibnamefont {Musch}}, \ and\ \bibinfo
  {author} {\bibfnamefont {A.}~\bibnamefont {Prokudin}},\ }\href {\doibase
  10.1007/JHEP10(2011)021} {\bibfield  {journal} {\bibinfo  {journal} {JHEP}\
  }\textbf {\bibinfo {volume} {10}},\ \bibinfo {pages} {021} (\bibinfo {year}
  {2011})},\ \Eprint {http://arxiv.org/abs/1107.5294} {arXiv:1107.5294
  [hep-ph]} \BibitemShut {NoStop}%
%%CITATION = ARXIV:1107.5294;%%
\bibitem [{\citenamefont {Scimemi}\ and\ \citenamefont
  {Vladimirov}(2016)}]{Scimemi:2016ffw}%
  \BibitemOpen
  \bibfield  {author} {\bibinfo {author} {\bibfnamefont {I.}~\bibnamefont
  {Scimemi}}\ and\ \bibinfo {author} {\bibfnamefont {A.}~\bibnamefont
  {Vladimirov}},\ }\href@noop {} {\  (\bibinfo {year} {2016})},\ \Eprint
  {http://arxiv.org/abs/1609.06047} {arXiv:1609.06047 [hep-ph]} \BibitemShut
  {NoStop}%
%%CITATION = ARXIV:1609.06047;%%
\bibitem [{\citenamefont {'t~Hooft}\ and\ \citenamefont
  {Veltman}(1972)}]{tHooft:1972tcz}%
  \BibitemOpen
  \bibfield  {author} {\bibinfo {author} {\bibfnamefont {G.}~\bibnamefont
  {'t~Hooft}}\ and\ \bibinfo {author} {\bibfnamefont {M.~J.~G.}\ \bibnamefont
  {Veltman}},\ }\href {\doibase 10.1016/0550-3213(72)90279-9} {\bibfield
  {journal} {\bibinfo  {journal} {Nucl. Phys.}\ }\textbf {\bibinfo {volume}
  {B44}},\ \bibinfo {pages} {189} (\bibinfo {year} {1972})}\BibitemShut
  {NoStop}%
%%CITATION = NUPHA,B44,189;%%
\bibitem [{\citenamefont {Breitenlohner}\ and\ \citenamefont
  {Maison}(1977)}]{Breitenlohner:1977hr}%
  \BibitemOpen
  \bibfield  {author} {\bibinfo {author} {\bibfnamefont {P.}~\bibnamefont
  {Breitenlohner}}\ and\ \bibinfo {author} {\bibfnamefont {D.}~\bibnamefont
  {Maison}},\ }\href {\doibase 10.1007/BF01609069} {\bibfield  {journal}
  {\bibinfo  {journal} {Commun. Math. Phys.}\ }\textbf {\bibinfo {volume}
  {52}},\ \bibinfo {pages} {11} (\bibinfo {year} {1977})}\BibitemShut {NoStop}%
%%CITATION = CMPHA,52,11;%%
\bibitem [{\citenamefont {Larin}(1993)}]{Larin:1993tq}%
  \BibitemOpen
  \bibfield  {author} {\bibinfo {author} {\bibfnamefont {S.~A.}\ \bibnamefont
  {Larin}},\ }\href {\doibase 10.1016/0370-2693(93)90053-K} {\bibfield
  {journal} {\bibinfo  {journal} {Phys. Lett.}\ }\textbf {\bibinfo {volume}
  {B303}},\ \bibinfo {pages} {113} (\bibinfo {year} {1993})},\ \Eprint
  {http://arxiv.org/abs/hep-ph/9302240} {arXiv:hep-ph/9302240 [hep-ph]}
  \BibitemShut {NoStop}%
%%CITATION = HEP-PH/9302240;%%
\bibitem [{\citenamefont {Larin}\ and\ \citenamefont
  {Vermaseren}(1991)}]{Larin:1991tj}%
  \BibitemOpen
  \bibfield  {author} {\bibinfo {author} {\bibfnamefont {S.~A.}\ \bibnamefont
  {Larin}}\ and\ \bibinfo {author} {\bibfnamefont {J.~A.~M.}\ \bibnamefont
  {Vermaseren}},\ }\href {\doibase 10.1016/0370-2693(91)90839-I} {\bibfield
  {journal} {\bibinfo  {journal} {Phys. Lett.}\ }\textbf {\bibinfo {volume}
  {B259}},\ \bibinfo {pages} {345} (\bibinfo {year} {1991})}\BibitemShut
  {NoStop}%
%%CITATION = PHLTA,B259,345;%%
\bibitem [{\citenamefont {Matiounine}\ \emph {et~al.}(1998)\citenamefont
  {Matiounine}, \citenamefont {Smith},\ and\ \citenamefont {van
  Neerven}}]{Matiounine:1998re}%
  \BibitemOpen
  \bibfield  {author} {\bibinfo {author} {\bibfnamefont {Y.}~\bibnamefont
  {Matiounine}}, \bibinfo {author} {\bibfnamefont {J.}~\bibnamefont {Smith}}, \
  and\ \bibinfo {author} {\bibfnamefont {W.~L.}\ \bibnamefont {van Neerven}},\
  }\href {\doibase 10.1103/PhysRevD.58.076002} {\bibfield  {journal} {\bibinfo
  {journal} {Phys. Rev.}\ }\textbf {\bibinfo {volume} {D58}},\ \bibinfo {pages}
  {076002} (\bibinfo {year} {1998})},\ \Eprint
  {http://arxiv.org/abs/hep-ph/9803439} {arXiv:hep-ph/9803439 [hep-ph]}
  \BibitemShut {NoStop}%
%%CITATION = HEP-PH/9803439;%%
\bibitem [{\citenamefont {Ravindran}\ \emph {et~al.}(2004)\citenamefont
  {Ravindran}, \citenamefont {Smith},\ and\ \citenamefont {van
  Neerven}}]{Ravindran:2003gi}%
  \BibitemOpen
  \bibfield  {author} {\bibinfo {author} {\bibfnamefont {V.}~\bibnamefont
  {Ravindran}}, \bibinfo {author} {\bibfnamefont {J.}~\bibnamefont {Smith}}, \
  and\ \bibinfo {author} {\bibfnamefont {W.~L.}\ \bibnamefont {van Neerven}},\
  }\href {\doibase 10.1016/j.nuclphysb.2004.01.001} {\bibfield  {journal}
  {\bibinfo  {journal} {Nucl. Phys.}\ }\textbf {\bibinfo {volume} {B682}},\
  \bibinfo {pages} {421} (\bibinfo {year} {2004})},\ \Eprint
  {http://arxiv.org/abs/hep-ph/0311304} {arXiv:hep-ph/0311304 [hep-ph]}
  \BibitemShut {NoStop}%
%%CITATION = HEP-PH/0311304;%%
\bibitem [{\citenamefont {Moch}\ \emph {et~al.}(2014)\citenamefont {Moch},
  \citenamefont {Vermaseren},\ and\ \citenamefont {Vogt}}]{Moch:2014sna}%
  \BibitemOpen
  \bibfield  {author} {\bibinfo {author} {\bibfnamefont {S.}~\bibnamefont
  {Moch}}, \bibinfo {author} {\bibfnamefont {J.~A.~M.}\ \bibnamefont
  {Vermaseren}}, \ and\ \bibinfo {author} {\bibfnamefont {A.}~\bibnamefont
  {Vogt}},\ }\href {\doibase 10.1016/j.nuclphysb.2014.10.016} {\bibfield
  {journal} {\bibinfo  {journal} {Nucl. Phys.}\ }\textbf {\bibinfo {volume}
  {B889}},\ \bibinfo {pages} {351} (\bibinfo {year} {2014})},\ \Eprint
  {http://arxiv.org/abs/1409.5131} {arXiv:1409.5131 [hep-ph]} \BibitemShut
  {NoStop}%
%%CITATION = ARXIV:1409.5131;%%
\bibitem [{\citenamefont {Vogelsang}(1998)}]{Vogelsang:1997ak}%
  \BibitemOpen
  \bibfield  {author} {\bibinfo {author} {\bibfnamefont {W.}~\bibnamefont
  {Vogelsang}},\ }\href {\doibase 10.1103/PhysRevD.57.1886} {\bibfield
  {journal} {\bibinfo  {journal} {Phys. Rev.}\ }\textbf {\bibinfo {volume}
  {D57}},\ \bibinfo {pages} {1886} (\bibinfo {year} {1998})},\ \Eprint
  {http://arxiv.org/abs/hep-ph/9706511} {arXiv:hep-ph/9706511 [hep-ph]}
  \BibitemShut {NoStop}%
%%CITATION = HEP-PH/9706511;%%
\bibitem [{\citenamefont {Mikhailov}\ and\ \citenamefont
  {Vladimirov}(2009)}]{Mikhailov:2008my}%
  \BibitemOpen
  \bibfield  {author} {\bibinfo {author} {\bibfnamefont {S.~V.}\ \bibnamefont
  {Mikhailov}}\ and\ \bibinfo {author} {\bibfnamefont {A.~A.}\ \bibnamefont
  {Vladimirov}},\ }\href {\doibase 10.1016/j.physletb.2008.11.051} {\bibfield
  {journal} {\bibinfo  {journal} {Phys. Lett.}\ }\textbf {\bibinfo {volume}
  {B671}},\ \bibinfo {pages} {111} (\bibinfo {year} {2009})},\ \Eprint
  {http://arxiv.org/abs/0810.1647} {arXiv:0810.1647 [hep-ph]} \BibitemShut
  {NoStop}%
%%CITATION = ARXIV:0810.1647;%%
\bibitem [{\citenamefont {Lefky}\ and\ \citenamefont
  {Prokudin}(2015)}]{Lefky:2014eia}%
  \BibitemOpen
  \bibfield  {author} {\bibinfo {author} {\bibfnamefont {C.}~\bibnamefont
  {Lefky}}\ and\ \bibinfo {author} {\bibfnamefont {A.}~\bibnamefont
  {Prokudin}},\ }\href {\doibase 10.1103/PhysRevD.91.034010} {\bibfield
  {journal} {\bibinfo  {journal} {Phys. Rev.}\ }\textbf {\bibinfo {volume}
  {D91}},\ \bibinfo {pages} {034010} (\bibinfo {year} {2015})},\ \Eprint
  {http://arxiv.org/abs/1411.0580} {arXiv:1411.0580 [hep-ph]} \BibitemShut
  {NoStop}%
%%CITATION = ARXIV:1411.0580;%%
\bibitem [{\citenamefont {Avakian}\ \emph {et~al.}(2008)\citenamefont
  {Avakian}, \citenamefont {Efremov}, \citenamefont {Schweitzer},\ and\
  \citenamefont {Yuan}}]{Avakian:2008dz}%
  \BibitemOpen
  \bibfield  {author} {\bibinfo {author} {\bibfnamefont {H.}~\bibnamefont
  {Avakian}}, \bibinfo {author} {\bibfnamefont {A.~V.}\ \bibnamefont
  {Efremov}}, \bibinfo {author} {\bibfnamefont {P.}~\bibnamefont {Schweitzer}},
  \ and\ \bibinfo {author} {\bibfnamefont {F.}~\bibnamefont {Yuan}},\ }\href
  {\doibase 10.1103/PhysRevD.78.114024} {\bibfield  {journal} {\bibinfo
  {journal} {Phys. Rev.}\ }\textbf {\bibinfo {volume} {D78}},\ \bibinfo {pages}
  {114024} (\bibinfo {year} {2008})},\ \Eprint {http://arxiv.org/abs/0805.3355}
  {arXiv:0805.3355 [hep-ph]} \BibitemShut {NoStop}%
%%CITATION = ARXIV:0805.3355;%%
\end{thebibliography}%



%%%%%%%%%%%%%%%%%%%%%%%%%%%%%%%%%
%%%%%%%%%%%%%%%%%%%%%%%%%%%%%%%%%
%%%%%%%%%%%%%%%%%%%%%%%%%%%%%%%%%

\end{document}